\documentclass[final,5p,times,twocolumn]{elsarticle}

\usepackage{graphicx}

\usepackage{amssymb}

\journal{Journal of Physics and Chemistry of Solids}

\begin{document}

\begin{frontmatter}



\title{Suppression of $T_c$ by Zn impurity in the electron-type LaFe$_{0.925-y}$Co$_{0.075}$Zn$_y$AsO system}


\author[1]{Yuke Li}
\author[1]{Jun Tong}
\author[1]{Qian Tao}
\author[1,2]{Guanghan Cao}
\author[1,2]{Zhuan Xu\corref{cor1}}\ead{zhuan@zju.edu.cn}

\address[1]{Department of Physics, Zhejiang University, Hangzhou 310027, P. R. China }
\address[2]{State Key Laboratory of Silicon Materials, Zhejiang University, Hangzhou 310027, P. R. China}

\cortext[cor1]{Corresponding author.}

\begin{abstract}
The effect of non-magnetic Zn impurity on superconductivity in
electron-type pnictide superconductor
LaFe$_{0.925-y}$Co$_{0.075}$Zn$_y$AsO is studied systematically.
The optimally doped LaFe$_{0.925}$Co$_{0.075}$AsO without Zn
impurity exhibits superconductivity at $T_c^{mid}$ of 13.2 K,
where $T_c^{mid}$ is defiend as the mid-point in the resistive
transition. In the presence of Zn impurity, the superconducting
transition temperature, $T_c^{mid}$, is severely suppressed. The
result is consistent with the theoretic prediction on the effect
of non-magnetic impurity in the scenario of $s_{\pm}$ pairing, but
it is in sharp contrast to the previous report on the effect of Zn
impurity in the F-doped systems. The possible interpretation of
the different effects of Zn impurity on superconductivity in
different systems is discussed.
\end{abstract}

\begin{keyword}
A. Superconductors \sep D. Defects
 \sep D. Superconductivity \sep
D. Transport Properties \sep D. Magnetic properties


\PACS 74.70.Dd \sep 74.20.Rp \sep 74.62.Dh

\end{keyword}

\end{frontmatter}



\section{Introduction}
\label{}

Great progresses have been made since the discovery of the first
FeAs-based pnictide superconductor LaFeAsO$_{1-x}$F$_x$ (La-1111
system) \cite{Kamihara08}. Similar to the high-temperature
superconducting (SC) cuprates, the parent compounds of the iron
pnictides are
antiferromagnet\cite{DaiPC-LaNeutron,BaoW-BaNeutron}.
Superconductivity can be induced by various chemical doping
methods, which introduces charge carriers and suppresses the
antiferromagnetic (AFM)
order\cite{Kamihara08,ChenXH-SmOF,WnagNL-CeOF,ZhaoZX-LnOD,WenHH-LaSr,WangC-GdTh,Co1,
Co2, Ni1, Ren-P}. It is generally agreed that magnetism could play
an important role to high-temperature superconductivity in both
cuprates and pnictides.  On the other hand, iron-based
superconductors also show different behaviors from the cuprates.
One of their major differences is the pairing symmetry. It has
been well established that the SC pairing in cuprates is of nodal
$d$-wave symmetry\cite{Tsuei1,Wollman}. In pnicitdes, the pairing
symmetry continues to be an important and outstanding issue. The
experimental evidences for full SC gaps in FeAs-based
superconductors are mainly from the measurements of angle-resolved
photoemission spectra (ARPES) \cite{ARPES}, Andreev tunnelling
spectra\cite{Andreev}, and penetration depth\cite{penetration}.
Within the full gap scenario, because of the multi-band Fermi
surfaces\cite{Singh}, the relative phases of the SC order
parameters in hole or electron pockets can be either positive
($s$-wave pairing) or negative ($s_{\pm}$-wave
pairing)\cite{Mazin,Kuroki,JPHu,Tes,LeeDH,ChenWQ}, depending on
the sign of the inter-Fermi pocket pair scattering amplitude or
their Josephson coupling. The $s_{\pm}$-wave pairing is appealing
with some experimental supports \cite{Tsuei2}. In addition, there
are also evidences for nodal SC gap in FeP-based
superconductors\cite{Moler}.

The effect of non-magnetic impurity is dependent on the pairing
symmetry and thus it can be used as a powerful probe to detect the
pairing symmetry for unconventional superconductors. Non-magnetic
impurity does not cause severe pair-breaking effect in a
conventional $s$-wave superconductor according to the Anderson's
theorem\cite{Anderson}. In the $d$-wave superconducting cuprates,
even a minimal amount of non-magnetic Zn doping can quench $T_c$
quickly\cite{Xiao}. In the $s_{\pm}$-wave state where the order
parameters in hole and electron pockets have opposite signs,
theoretical studies\cite{Bang,Kontani} have predicted that
non-magnetic impurities like Zn can severely suppress the SC
transition temperature $T_c$, similar to the effect in high-$T_c$
cuprates with $d$-wave pairing. Zinc element has a stable $d^{10}$
configuration in the alloy \cite{Singh2009}, and can serve as the
best non-magnetic impurity for this study. However, there is a
seemingly discrepancy between our early data in
LaFeAsO$_{0.9}$F$_{0.1}$, where $T_c$ is robust to
Zn-impurity\cite{LaOZnAs}, and a following report showed a severe
suppression of $T_c$ due to Zn-impurity in the oxygen-deficient
LaFeAsO$_{1-\delta}$ samples\cite{nims}. Further studies found
that the effect of Zn impurity on $T_c$ is strongly dependent on
the charge doping level, i.e., $T_c$ increases with Zn content in
the under-doped regime, remains unchanged in the optimally doped
regime, and is severely suppressed in the over-doped regime in the
LaFe$_{1-y}$Zn$_y$AsO$_{1-x}$F$_x$ system\cite{Li-Zn2}. This
finding reconciles the contradiction in the previous reports of Zn
impurity effect and strongly suggests that a switch of pairing
symmetry from an impurity-insensitive state to an
impurity-sensitive state is possible.

In this paper, we study the effect of Zn impurity on
superconductivity in an optimally Co-doped
LaFe$_{0.925-y}$Co$_{0.075}$Zn$_y$AsO system. We found that $T_c$
is severely suppressed with addition of Zn impurity although the
system is in the optimally doped regime. This result is in
contrast to the effect of Zn impurity in optimally F-doped La-1111
system\cite{LaOZnAs}. The possible interpretations regarding to
the SC pairing state are discussed.

\begin{figure}[htbp]
 \includegraphics[width=7cm]{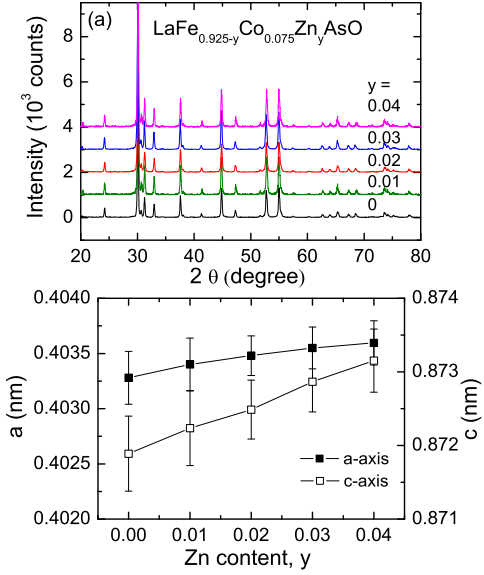}
 \caption{(a) Powder X-ray diffraction
patterns of LaFe$_{0.925-y}$Co$_{0.075}$Zn$_{y}$AsO samples. (b)
Variations of lattice parameters as a function of Zn
content.}\label{fig1}
 \end{figure}

\section{Experimental}
Polycrystalline samples of LaFe$_{0.925-y}$Co$_{0.075}$Zn$_y$AsO
($y$ = 0, 0.01, 0.02, 0.03, 0.04) were synthesized by solid state
reaction method. Details on the sample preparation can be found in
the previous report\cite{LaOZnAs}. The phase purity of the samples
was investigated by powder X-ray diffraction (XRD) using a
D/Max-rA diffractometer with Cu-K$_{\alpha}$ radiation and a
graphite monochromator. Lattice parameters were calculated by a
least-squares fit using at least 20 XRD peaks.

The electrical resistivity was measured on bar-shaped samples
using a standard four-probe method. The measurements of resistance
and Hall effect were performed on a Quantum Design Physical
Property Measurement System (PPMS-9). DC magnetization were
measured on a Quantum Design Magnetic Property Measurement System
(MPMS-5).

\begin{figure}[htbp]
 \includegraphics[width=7cm]{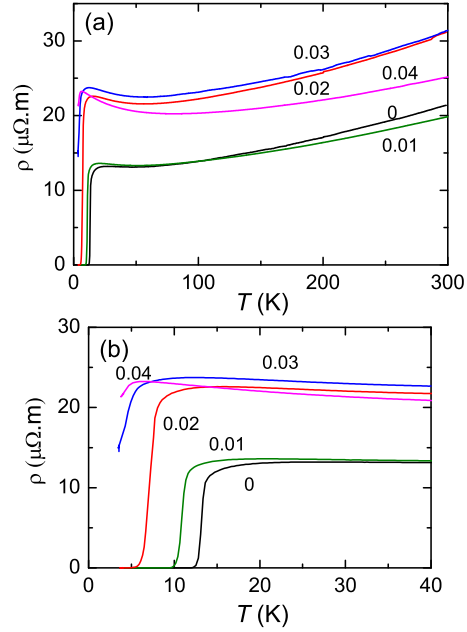}
 \caption{(a) Temperature dependence of
resistivity ($\rho$) for LaFe$_{0.925-y}$Co$_{0.075}$Zn$_{y}$AsO
samples. (b) Enlarged plots of $\rho$ versus $T$ around the
transition temperature.}\label{fig2}
 \end{figure}

\section{Results and Discussion}

Figure 1(a) shows the XRD patterns of the
LaFe$_{0.925-y}$Co$_{0.075}$Zn$_y$AsO samples, and Figure 1(b)
shows the variations of lattice constants with Zn content. All the
XRD peaks can be well indexed with the tetragonal ZrCuSiAs-type
structure, indicating that all the samples are single phase
without foreign phases. Both the $a$-axis and $c$-axis increase
slightly with the Zn content, consistent with the fact that the
lattice constants of LaZnAsO are larger than those of
LaFeAsO\cite{RZnAsO}. These results indicate that Zn has
successfully substituted partial Fe ions in the lattice.

Figure 2 shows the temperature dependence of resistivity for the
LaFe$_{0.925-y}$Co$_{0.075}$Zn$_{y}$AsO samples. The resistivity
around the SC transition is enlarged as shown in the Figure 2(b).
According to our previous study, the LaFe$_{0.925}$Co$_{0.075}$AsO
sample without Zn doping is optimally doped electron-type
superconductor\cite{Co2}. The SC transition temperature
$T^{mid}_c$ (defined as the midpoint in the resistive transition)
is 13.2 K. With the presence of Zn impurity, $T^{mid}_c$ is
suppressed quickly. Meanwhile the normal state resistivity
increases with Zn doping, and there is an obvious upturn at low
temperatures just above $T_c$. 4\% substitution of Fe by Zn has
already quenched superconductivity ($T_c$ at least below 2 K).
Recall that superconductivity is found to be robust to the Zn
impurity in the optimally F-doped electron-type LaFeAsO system,
and also in the optimally K-doped hole-type (Ba,K)Fe$_2$As$_2$,
this result is quite surprising and suggests a crucial difference
in the Co-doped pnictides where the chemical doping occurs
directly in the conducting FeAs layers.

Figure 3 shows the temperature dependence of d.c. magnetic
susceptibility measured under $H$ of 10 Oe. Without Zn doping, the
7.5\% Co doping makes the system into optimally-doping state with
a maximum $T^{mid}_{c}$ of 13.2 K. The sharp transition in the
magnetic susceptibility below $T_{c}$ suggests bulk
superconductivity and high homogeneity, consistent with the
previous report\cite{Co2}. Actually the volume fraction of
superconducting magnetic shielding even exceeds 100 \% for this
samples. The demagnetizing factor $N$ has not been taken into
account, thus the volume fraction could be much over-estimated.
With addition of Zn impurity, the transition in magnetic
susceptibility moves to lower temperatures, concsistent with the
resistivity data. Meanwhile, the volume fraction of
superconducting magnetic shielding becomes smaller.

Figure 4 shows temperature dependence of Hall coefficient $R_{H}$
for LaFe$_{0.925-y}$Co$_{0.075}$Zn$_y$AsO samples. For the
optimally doped samples without Zn impurity, $R_H$ shows a sharp
transition when it enter into superconducting state. There is a
large negative peak just below $T^{mid}_{c}$, and then $R_H$ goes
to zero with further deceasing temperature. Such a negative peak
in $R_H$ below $T_c$ is often observed in high-quality high-$T_c$
cuprates and it is ascribed to the contribution from the motion of
vortices\cite{Ong}. The presence of Zn impurity diminishes this
peak. Meanwhile, the high-temperature $R_{H}$ remains almost
unchanged with Zn doping, indicating that Zn doping does not
change the charge carrier density. This is consistent with the
band calculation result which predicts that energy level of Zn
3$d$ electrons are deep below the Fermi level\cite{Singh2009}.
Unlike Co doping, Zn doping does not induce extra mobile electrons
into the conducting FeAs layers.

 \begin{figure}[htbp]
 \includegraphics[width=7cm]{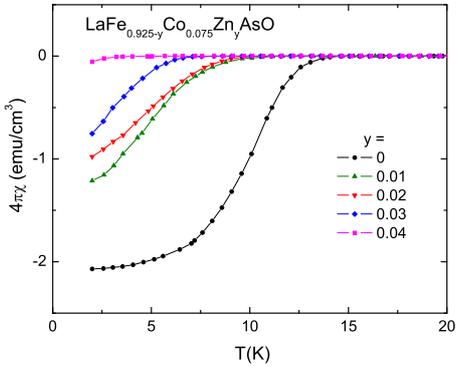}
 \caption{Temperature dependence of magnetic susceptibility of LaFe$_{0.925-y}$Co$_{0.075}$Zn$_{y}$AsO samples measured under $H$ of 10 Oe. The data were taken under the zero-field cooling (ZFC) protocol.}\label{fig3}
 \end{figure}

\begin{figure}[htbp]
 \includegraphics[width=7cm]{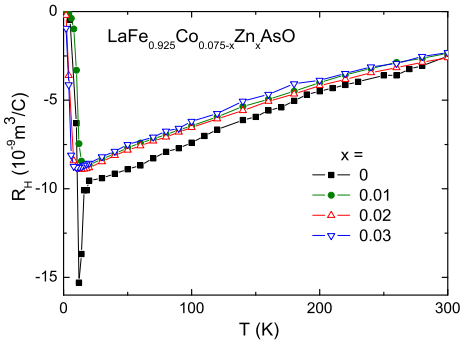}
 \caption{Temperature dependence of Hall coefficient, $R_H$, of the LaFe$_{0.925-y}$Co$_{0.075}$Zn$_{y}$AsO samples measured under $H$ of 50000 Oe.}\label{fig4}
 \end{figure}

We plot $T_c$ versus Zn content ($x$) of
LaFe$_{0.925-y}$Co$_{0.075}$Zn$_y$AsO and $T_c$ versus Co content
($x$) of LaFe$_{1-x}$Co$_{x}$AsO in Figure 5 to compare the effect
of Zn impurity with that of Co doping. The data of $T_c$ with $x$
of LaFe$_{1-x}$Co$_{x}$AsO are taken from our previous
report\cite{Co2}. As we pointed out hereinbefore, the maximum
$T_c$ is reached at the optimally doping level $x$ = 0.075 for
this Co-doped electron-type La-1111 superconductor. When Co
content is greater than 0.075, $T_c$ decreases gradually with a
rate (d$T_c/dx$) of about -1.0 K per 1\% of Co doping due to
over-doping of charge carriers. However, it is obvious that the
decreasing rate of $T_c$ with $y$ (Zn content) is much larger,
about -3.0 K per 1\% of Zn impurity, implied that there is
fundamentally different mechanism of Zn doping. The result also
supports our conclusion that the Zn doping does not induce extra
mobile electrons to the conducting FeAs layers, whereas Co doping
does. Actually the resistivity value increases quickly with
increasing Zn content, and the upturn in resistivity at low
temperature also becomes more obvious, implying that the Zn
impurity is a very effective impurity scattering center and it
causes  severe SC pairing breaking effect in the Co-doped
electron-type La-1111 systems. This finding is in sharp contrast
to the previous report on the effect of Zn impurity in the
optimally F-doped La-1111 system where $T_c$ is robust to the Zn
impurity\cite{LaOZnAs}. Nevertheless, a severe suppression of
$T_c$ by Zn impurity is indeed observed in over-doped La-1111
systems \cite{nims,Li-Zn2}. In previous study, a switch of paring
state from impurity-insensitive state (such as usual $s$-wave) to
impurity-sensitive state (such as $d$-wave or $s_{\pm}$-wave) has
been proposed when the F doping level is increased. In the
Co-doped electron-type systems, the pairing state could become
impurity-sensitive (e.g., $d$-wave or $s_{\pm}$-wave) even in the
optimally doped level. Such a change in pairing state with
different chemical doping (especially on different lattice site)
strongly suggests that the even a very subtle change in the
electronic structure could result in a different pairing state,
i.e. the pairing state could be dependent on the chemical doping.
This proposal is also supported by the previous reports on the
measurements of SC gap. A nodal gap has been observed for the
P-doped BaFe$_2$As$_2$ system\cite{P122} and also in over-doped
KFe$_2$As$_2$ system\cite{K122} while full gaps have been observed
in the optimally doped BaFe$_2$As$_2$ system\cite{ARPES}.
Furthermore, we should also point out another possibility that the
nature of Co doping could be disturbed by Zn impurity since both
are substituted on the FeAs layer. Furthermore studies are
required to clarify the different mechanisms of chemical doping.

\begin{figure}[htbp]
 \includegraphics[width=7cm]{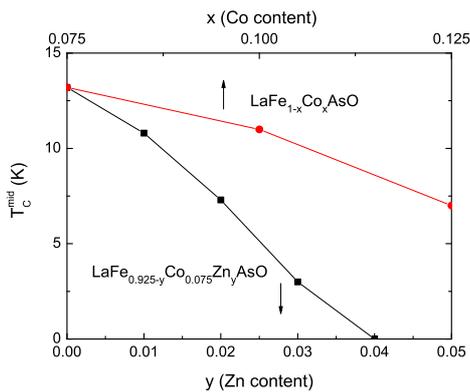}
 \caption{Variation of $T^{mid}_c$ with Zn content for the LaFe$_{0.925-y}$Co$_{0.075}$Zn$_{y}$AsO system. The $T^{mid}_c$ versus Co content in the over-doped regime of the LaFe$_{1-x}$Co$_{x}$AsO system is also shown for comparison. The $T^{mid}_c$ data for the LaFe$_{1-x}$Co$_x$AsO system are taken from Ref.\cite{Co2}.}\label{fig5}
 \end{figure}

\section{Conclusion}

In summary, we find a severe suppression of superconductivity by
non-magnetic Zn impurity in the Co-doped electron-type La 1111
system LaFe$_{0.925-y}$Co$_{0.075}$Zn$_y$AsO. This finding is in
contrast to the previous studies of Zn impurity effect in
optimally F-doped system, although it is consistent with the
theoretic prediction on the effect of non-magnetic impurity in the
scenario of $s_{\pm}$ pairing. These results on the effects of Zn
impurity strongly suggest that the pairing state could be
different in the systems with different chemical doping. The
difference in the electronic structure due to different chemical
doping should be taken into account by the theoretical models.

We gratefully acknowledge the support from National Science
Foundation of China (Grant Nos 10634030 and 10931160425), and the
National Basic Research Program of China (Grant Nos 2007CB925001
and 2011CB605903).



\end{document}